\documentclass[aps,preprint,showpacs]{revtex4}
\usepackage{graphicx}
\begin{document}
\title{Versatility of field theory motivated nuclear effective  \\
Lagrangian approach}
\author{P. Arumugam} %\email{aru@iopb.res.in}
\author{B. K. Sharma} %\email{bharat@iopb.res.in}
\author{P. K. Sahu} %\email{pradip@iopb.res.in}
\author{S. K. Patra} %\email{patra@iopb.res.in}
\affiliation{Institute of Physics, Sachivalaya Marg, Bhubaneswar -
751 005, India.}

\author{Tapas Sil}
\author{M. Centelles}
\author{X. Vi\~nas}
\affiliation{\it Departament d'Estructura i Constituents de la Mat\`eria,
Facultat de F\'{\i}sica,
\\
Universitat de Barcelona,
Diagonal {\sl 647}, {\sl 08028} Barcelona, Spain}

\begin{abstract}
We analyze the results for infinite nuclear and neutron matter using
the standard relativistic mean field model and its recent effective
field theory motivated generalization. For the first time, we show
quantitatively that the inclusion in the effective theory of vector
meson self-interactions and scalar-vector cross-interactions explains
naturally the recent experimental observations of the softness of the
nuclear equation of state, without losing the advantages of the
standard relativistic model for finite nuclei.
\end{abstract}
             

\maketitle

\section{Introduction}

In the quest for a unified model describing both finite nuclei and
nuclear matter properties, the relativistic mean field (RMF) approach
to quantum hadrodynamics \cite{ser86} has become a very popular tool.
The original linear $\sigma$-$\omega$ model of Walecka \cite{ser86}
was complemented on an \textit{empirical} basis with cubic and quartic
non-linearities \cite{Bo77} of the $\sigma$ meson to have proper
results. We shall refer to this approach as standard RMF model. In
recent years novel approaches based on modern concepts of effective
field theory (EFT) and density functional theory (DFT) for hadrons, in
different implementations, have been developed for the relativistic
nuclear many-body problem
\cite{Fu96,Se97,Mu96,Fu00,kaiser02,kaiser02b}. One of the goals is to
overcome the \textit{ad hoc} nature of the previous models and the
presence of uncontrolled approximations. Here we address the chiral
effective Lagrangian model proposed by Furnstahl, Serot and Tang
\cite{Fu96}, whose mean field treatment is hereafter called E-RMF
model. A nuclear EFT contains all the non-renormalizable couplings
consistent with the underlying symmetries of QCD\@. The effective
Lagrangian is expanded in powers of the fields and their derivatives,
and the possible terms and their importance is systematically
categorized \cite{Fu96,Se97,Mu96,Fu00}. None of the couplings present
at a given order can be arbitrarily dropped out without a symmetry
argument. 

The free parameters of the E-RMF Lagrangian have been optimized by
fitting to the ground-state properties of a few doubly-magic nuclei
\cite{Fu96}. The truncation of the effective Lagrangian at the first
lower orders is validated by the fact that when one includes up to
fourth order terms in the expansion, the E-RMF fits (parameter sets G1
and G2 determined in Ref.\ \cite{Fu96}) display naturalness and the
results are not dominated by the last terms retained
\cite{Fu96,Se97,Mu96,Fu00,Es99,estal1,estal2,huertas}.

Very recently, the flow of matter in heavy-ion collisions has been
analyzed to determine experimentally the pressures attained at
densities ranging from two to five times the saturation density of
nuclear matter \cite{Sci02}. Nuclear collisions in man-made
laboratories can nowadays compress nuclear matter to such extreme
densities, which occur in Nature within neutron stars and
core-collapse supernovae. Nuclear models of various sorts that are
tuned to the known properties of terrestrial nuclei at normal
densities or to nucleon-nucleon (NN) scattering data are commonly
extrapolated for predictions of those very dense systems. The recent
breakthrough achieved in the experimental determination of the
equation of state (EOS) of high-density matter \cite{Sci02} motivates
us to study the applicability of the RMF and E-RMF models to those
conditions. The performance of the E-RMF approach for finite nuclei
has already been demonstrated in a series of previous works
\cite{Fu96,estal1,estal2,huertas,SIL}.

\section{Discussion}

The energy density functional of the E-RMF model is written as
\cite{Se97,Fu96}: 
%
%% Equation 1
\begin{eqnarray}
& & {\cal E}(\textbf{r}) =
\nonumber \\[3mm]
& & \sum_\alpha \varphi_\alpha^\dagger
\Bigg\{ -i \mbox{\boldmath$\alpha$} \!\cdot\!
\mbox{\boldmath$\nabla$} + \beta (M - \Phi) + W +
\frac{1}{2}\tau_3 R + \frac{1+\tau_3}{2} A - \frac{i}{2M} \beta
\mbox{\boldmath$\alpha$}\!\cdot\! \mbox{\boldmath$\nabla$} 
\left( f_v W + \frac{1}{2} f_\rho\tau_3 R + \lambda A \right)
\nonumber \\[3mm]
& & + \frac{1}{2M^2}\left (\beta_s + \beta_v \tau_3 \right )
\Delta A \Bigg\} \varphi_\alpha \null + \left ( \frac{1}{2} +
\frac{\kappa_3}{3!}\frac{\Phi}{M} +
\frac{\kappa_4}{4!}\frac{\Phi^2}{M^2}\right )
\frac{m_{s}^2}{g_{s}^2} \Phi^2  - \frac{\zeta_0}{4!} \frac{1}{
g_{v}^2 } W^4
\nonumber \\[3mm]
& & \null + \frac{1}{2g_{s}^2}\left( 1 +
\alpha_1\frac{\Phi}{M}\right) \left( \mbox{\boldmath
$\nabla$}\Phi\right)^2 - \frac{1}{2g_{v}^2}\left( 1
+\alpha_2\frac{\Phi}{M}\right) \left( \mbox{\boldmath $\nabla$} W
\right)^2 \null - \frac{1}{2}\left(1 + \eta_1 \frac{\Phi}{M} +
\frac{\eta_2}{2} \frac{\Phi^2 }{M^2} \right)
\frac{{m_{v}}^2}{{g_{v}}^2} W^2
\nonumber \\[3mm]
& &  - \frac{1}{2g_\rho^2} \left( \mbox{\boldmath $\nabla$}
R\right)^2 - \frac{1}{2} \left( 1 + \eta_\rho \frac{\Phi}{M}
\right) \frac{m_\rho^2}{g_\rho^2} R^2 \null - \frac{1}{2e^2}\left(
\mbox{\boldmath $\nabla$} A\right)^2 + \frac{1}{3g_\gamma g_{v}}A
\Delta W + \frac{1}{g_\gamma g_\rho}A \Delta R \, .
\label{eq1}
\end{eqnarray}
The energy density (\ref{eq1}) contains tensor couplings and
scalar-vector and vector-vector meson interactions in addition to the
standard scalar self-interactions $\kappa_3$ and $\kappa_4$. One
interprets the E-RMF formalism as a covariant formulation of DFT\@.
That is, the mean field model approximates the exact (but unknown)
energy functional of the many-nucleon system, which includes all
higher-order correlations, by expanding it in powers of the meson
fields. The latter play the role of auxiliary Kohn-Sham potentials.
The unresolved dynamics beyond mean field is encoded in the fitted
coupling constants of the effective theory and the introduction of new
interaction terms aims at an improved representation of the nuclear
energy functional. Further insight into the concepts of the E-RMF
model can be gained from Refs.\ \cite{Fu96,Se97,Mu96,Fu00}.

For uniform nuclear matter all of the terms with gradients in the
energy density ${\cal E}$ vanish and the nucleon
density is given by $\rho= \frac{\gamma}{6{\pi^2}}{k_F^3}$ 
($\gamma$=4 for symmetric nuclear matter, $\gamma$=2 for neutron
matter, and $k_F$ is the Fermi momentum). The pressure $P$ follows
from the derivative of the energy density with respect to the nucleon
density: $P = \rho^2 \frac{\partial}{\partial\rho} ({\cal E}/\rho)$.

\begin{figure}
\includegraphics[width=0.65\textwidth,clip=true]{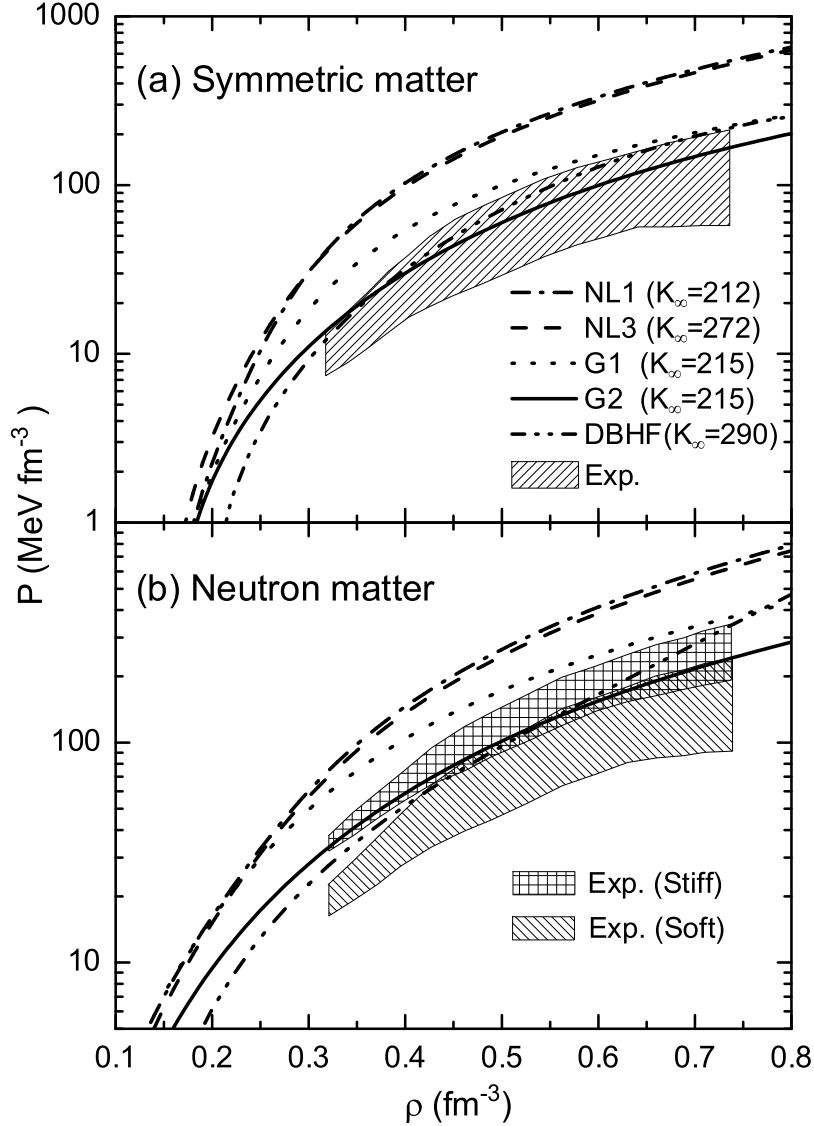}
\caption{\label{fig1} (a) Zero temperature EOS for symmetric nuclear
matter. The shaded area represents the region consistent with the
experimental data \cite{Sci02}. The nuclear matter incompressibility
of each interaction is expressed in MeV\@. (b) Zero temperature EOS
for neutron matter. The upper and lower shaded areas depict the
regions compatible with experiment after inclusion of the pressure
from asymmetry terms with, respectively, strong and weak density
dependences \cite{Sci02}.}
\end{figure}

The recent experimental observations \cite{Sci02,PRL01} rule out any
strongly repulsive nuclear EOS\@. The constraints on the EOS of
symmetric nuclear matter at zero temperature, derived experimentally
by analyzing elliptic and transverse flow observables in nuclear
collisions, are shown in Fig.\ 1a along with the predictions from
different RMF and \mbox{E-RMF} sets. We can see that the calculations
of dense matter based on NL1 \cite{reinhard86} and NL3 \cite{Lal97}
deviate drastically from experiment, while the E-RMF calculations with
G1 (to some extent) and G2 (to a better extent) agree more with the
allowed region. A similar situation prevails in the EOS of neutron
matter (Fig.\ 1b). The figures show that conventional sets such as NL1
and NL3 are not well suited for describing the EOS of dense matter,
principally due to the fact that without the additional self- and
cross-interactions the density dependence of the mesonic mean fields
is too poor \cite{estal1,Sug94}. Notice that although the
incompressibility of NL1 is $K_{\infty}=212$ MeV, which is within the
empirical limit, the resulting EOS is stiff at higher densities and
does not follow the experimental trend with increasing density. On the
other hand, the E-RMF calculations explain better the situation
without any forced changes in the fitted parameters or the formalism,
and the consistency of G2 with experiment is outstanding. As depicted
in Fig.\ 1, the microscopic Dirac-Brueckner-Hartree-Fock (DBHF) theory
\cite{Broc90}, which starts from the NN interaction in free space,
also suggests an EOS for dense matter much softer than in the usual
RMF approach. One notices that the DBHF results, for both nuclear and
neutron matter, lie within the regions of the EOS compatible with the
boundaries extracted from experiment.

The success of G2 in nuclear matter is possible owing to the bulk
couplings $\zeta_0$, $\eta_1$, and $\eta_2$ which confer an extra
density dependence to the scalar and vector self-energies. If using
the functional form (\ref{eq1}) one attemps to get a soft EOS for
dense matter with vanishing $\eta_1$ and $\eta_2$, for realistic
saturation conditions, this tends to enforce large unnatural values of
$\zeta_0$. If with $\eta_1= \eta_2= 0$ one keeps $\zeta_0$ at
acceptable values, then $\kappa_4$ tends to become negative. With
inclusion of the additional parameters $\zeta_0$, $\eta_1$ and
$\eta_2$ one can agree better with experiment, have $\kappa_4>0$ and a
not very large $\zeta_0$ value, as in the case of G2. The experimental
data of Fig.~1 also are explained reasonably well \cite{Sci02} by the
calculations of Akmal \textit{et al} \cite{Akmal} which employ the
Argonne $v_{18}$ interaction, though the use of this interaction for
finite nuclei is still limited. It will be worth exploring the
situation for relativistic models that taking advantage of
density-dependent coupling vertices are consistent with the DBHF
theory \cite{typel03}.

In Table I we present the results for the mass and radius of a neutron
star. The structure of the neutron star is determined through the
Tolman-Oppenheimer-Volkov equation, which follows from hydrostatic
equilibrium in strong gravitational fields and general relativity
\cite{Wein72}, combined with the EOS for $P(\rho)$. One sees that the
neutron star properties computed with G2 are very much in accord with
the available experimental information. The critical temperatures for
the liquid-gas phase transition from our calculations of nuclear
matter at finite temperature are also displayed in Table I\@. These
values are governed by the EOS around saturation where the E-RMF sets
produce similar results to NL3. Hence, one expects that in the finite
nuclei regime G1 and G2 yield results on a par with the celebrated NL3
parametrization. That this is indeed the case has been proven in
previous works and therefore we shall not attempt to present more
results in this respect here. We rather refer the reader to the
literature \cite{Fu96,Es99,estal1,estal2,huertas,SIL} where it has
been shown that the bulk and single-particle properties of finite
nuclei are well described by the \mbox{E-RMF} model with a similar
quality to the standard RMF parametrizations, not only for stable
isotopes but also for exotic isospin-rich nuclei far from the valley
of $\beta$ stability.

\begin{table}
\caption{The radius $R$ (in km) and the mass ratio $M/M_\odot$ of a
neutron star, and the critical temperature $T_c$ (in MeV) for the
liquid-gas phase transition in nuclear matter and in asymmetric matter
with asymmetry $\alpha=(\rho_n-\rho_p)/(\rho_n+\rho_p)=0.2$.}
\begin{ruledtabular} \begin{tabular}{llrrrrr}
%\hline \hline
&                 &   NL1    &    NL3    &    G1     &    G2     &    Exp. \\ \hline
&  $R$            &  16.94   &   15.47   &   13.88   &   10.08   &    10--12   \\
&  $M/M_\odot$    &   2.93   &    2.78   &    2.15   &   2.05    &    1.5--2.5     \\
\hline
&  $T_c$ (Sym.)   &  13.5    &   14.3    &    14.3   &   14.2    &    13--20 \\
&  $T_c$ (Asym.)  &  12.9    &   13.7    &    13.8   &   13.7    &                 \\
\end{tabular}
\end{ruledtabular}
\end{table}

In choosing between the parameter sets G1 and G2 for further
calculations we prefer G2. It is worth noting that G2 presents a
positive value of the $\Phi^4$ coupling constant ($\kappa_4$), as
opposed to G1 and to most of the successful RMF parametrizations, such
as NL3. Though the energy spectrum strictly has no lower bound with a
negative $\kappa_4$ \cite{Baym60}, such negative value is necessary in
the standard RMF model to get the results
closer to experiment. On the other hand, to have a positive $\kappa_4$
it is not indispensable to make two parameter sets (one for light and
one for heavy nuclei) as in Ref.\ \cite{Sug94}. 

\section{Summary}

In conclusion, we have shown for the first time that the E-RMF
parameter sets give a soft EOS both around saturation and at high
densities which is consistent with the measurements of kaon production
\cite{PRL01} and the flow of matter \cite{Sci02} in energetic
heavy-ion collisions, and with the observed neutron star masses and
radii. With the systematic inclusion of new interaction terms under
the guidance of EFT techniques, and without forcing any change of the
parameters initially determined from a few magic nuclei, the E-RMF
calculations with the G2 set explain finite nuclei and nuclear matter
in a unified way with a commendable level of accuracy in both the
cases. The authors are not aware that it has been possible to prove
this for other effective nuclear models. At present, the E-RMF
approach can be considered as a salient step towards a unified theory
for finite nuclei as well as for infinite nuclear matter and
highlights the potential of the application of effective field theory
formulations for nuclear structure studies.

\acknowledgments{Work partially supported by grants BFM2002-01868
(DGI, Spain, and FEDER) and 2001SGR-00064 (DGR, Catalonia).
T. S. also thanks the Spanish Education Ministry grant SB2000-0411 for
financial support.}

\end{document}